\documentclass[12pt,preprint]{aastex}

\def\vec#1{{\bf #1}}
\def\balpha {\mbox{\boldmath $\alpha$}}
\def\bbeta {\mbox{\boldmath $\beta$}}
\def\deg{\ifmmode^\circ\else$^\circ$\fi}
\def\Fscr{\ifmmode{\mathcal{F}}\else$\mathcal{F}$\fi}
\def\Ascr{\ifmmode{\mathcal{A}}\else$\mathcal{A}$\fi}

\makeatletter
\slugcomment{}
\shorttitle{Calibrated Throughput Measurements}
\shortauthors{C.W.~Stubbs and J.L.~Tonry}

\makeatother
\begin{document}

\title{Toward 1\% Photometry: End-to-end Calibration of Astronomical
Telescopes and Detectors.}

\author{Christopher W. Stubbs (1)}

\altaffiltext{1}{Department of Physics and Harvard-Smithsonian Center for 
Astrophysics, Harvard University}

\author{John L. Tonry (2)}
\altaffiltext{2}{Institute for Astronomy, University of Hawaii}

\begin{abstract}

We review the systematic uncertainties that have plagued attempts to
obtain high precision and high accuracy from ground-based photometric
measurements using CCDs.  We identify two main challenges in breaking through the
1\% precision barrier: 1) fully characterizing atmospheric
transmission, along the instrument's line of sight, and 2) properly
identifying, measuring and removing instrumental artifacts.  We
discuss approximations and limitations inherent in the present
methodology, and we estimate their contributions to systematic
photometric uncertainties.  We propose an alternative conceptual
scheme for the relative calibration of astronomical apparatus: the
availability of calibrated detectors whose relative spectral
sensitivity is known to better than one part in $10^3$ opens up the
possibility of {\it in situ} relative throughput measurements,
normalized to a precision calibrated detector, using a stable but
uncalibrated narrowband light source.  An implementation scheme is
outlined, which exploits the availability of tunable lasers to map out
the relative wavelength response of an imaging system, using a
flatfield screen and a calibrated reference photodiode. The merits and
limitations of this scheme are discussed.  In tandem with careful
measurements of atmospheric transmission, this approach could
potentially lead to reliable ground-based photometry with fractional
uncertainties below the percent level.

\end{abstract}


\section{Introduction and motivation}
It has proven difficult to achieve ground-based flux measurements with
fractional uncertainties in accuracy of less than a few percent, using CCDs. 
This is much
poorer than naive Poisson statistics would permit. Most photometry
codes add a fractional flux uncertainty of a few percent in quadrature
with the uncertainty from photon statistics, in recognition of various
sources of systematic error \cite{Stetson87}, \cite{Schechter93}.  The
Sloan Digital 
Sky Survey (SDSS), for example, quotes \cite{SDSS02} a systematic
photometric zeropoint rms uncertainty of 2\% across the survey
fields. Saha et al.~\cite{Saha05} compare precision photometry from multiple sources, and find systematic discrepancies at the 
2-5\% level. They attribute these discrepancies to passband 
differences and 

On the other hand, differential photometric measurements of
sources with similar spectral energy distributions achieve
\cite{Hartman04} \cite{Tonry05} precise differential flux measurements
at the millimagnitude
level, as long as the multiple sources can be captured in a single
frame of the imaging system. Laboratory tests of the stability of CCD
detectors indicate that they are capable of achieving Poisson--limited
differential measurements at the level of one part in $10^5$
\cite{Robinson95}. The intrinsic stability of CCDs has led to
ambitious plans for high resolution differential photometry from space
\cite{Borucki03}.

While very impressive  {\em relative} flux measurements can now be made
between objects that reside in a single exposure, determining the 
flux ratios between objects in different regions of the sky is 
much more difficult. This is one indicator that the temporal and 
directional variation in the optical transmission of the atmosphere are
significant limitations to precision photometry. 

It should be stressed that the normal procedures for photometric
measurement are quite satisfactory for the purpose for which they were
designed: measuring stars at the few percent level.  Since 25\% of
stars are variable at the 1\%~rms level or more (and 50\% at the 0.5\%
level) \cite{Tonry05}, and since stellar colors are often corrupted at
the percent level by dust, there has been little impetus to do better.

A number of forefront science issues demand better performance from
ground-based systems. The surge of interest in extrasolar planetary
systems motivates searches for planetary transits, often via a subtle
photometric signal. On the cosmological scale, using supernovae to
probe the expansion history of the Universe requires making
K-corrections \cite{Peacock98} that rely upon detailed knowledge of
the relative 
instrumental response as a function of wavelength.  These
considerations motivate taking a close look at how astronomical flux
measurements are made, how the apparatus is calibrated, and how the
data are analyzed. This paper will explore how we might break through
the photometric precision barrier at the percent level, for
ground--based CCD flux measurements.
 
In the sections that follow we will argue the following points:

\begin{enumerate}

\item{} ``Flatfielding'' CCD images by dividing the flux
detected in each pixel by a passband dependent sensitivity 
array is susceptible to systematic 
uncertainties due to differences in the spectral energy distributions 
of the flatfield, the sky, and the source. 

\item{} Numerous approximations commonly made in the reduction of 
broadband images are likely to fail in attempts to measure
flux with fractional uncertainty below the 1\% level. 

\item{} It is possible and desirable to measure the relative
throughput of the entire imaging system in situ, in which the 
wavelength--dependent throughput of the mirrors, corrector optics,
filter, and detector are determined using a tunable laser as a high
brightness monochromatic source and a calibrated photodiode as a
detector.

\item{} There is considerable merit in obtaining ongoing measurements
of atmospheric transmission over wavelength, using a dedicated
spectrophotometric instrument in conjunction with modern atmospheric
models.

\item{} Future imaging systems should be designed for ease of calibration
and suppression of sources of systematic error. One obvious example is
the inclusion of an independent determination of shutter timing. 

\end{enumerate}

Numerous authors have explored the issues that pertain to the
calibration of photometric measurements, including \cite{Bessell99},
\cite{Bohlin96}, \cite{Castelli94}, \cite{Colina94},
\cite{Fabregat96}, \cite{Fukugita96}, \cite{Hamuy92}, \cite{Hamuy94}, 
\cite{Hayes75}, \cite{Hayes75b}, \cite{Landolt92}, \cite{Magnier04}, 
\cite{Megessier95}, \cite{Oke83}, \cite{Saha05},  and others. Our focus here is less on the use of
celestial calibrators than on exploiting the availability of
calibrated silicon photodiodes, for characterizing astronomical
apparatus, and on the diverse contributions to {\it systematic}
uncertainty in astronomical flux measurements.

Current practice in astronomical flux measurement makes an estimate of
throughput for each passband, using filter transmission data obtained
from benchtop measurements, throughput estimates for optical
components, and a detector quantum efficiency curve. Broadband
measurements of standards are then used to ``tweak'' the system throughput
function to force agreement between measurements and synthetic
photometry \cite{Bessell05}. Unfortunately the integral measurement of
flux over a broad passband does not uniquely determine what passband
modification is to be made.

The metrology chain that underpins most astronomical flux measurements
dates back to an era when laboratory blackbody sources of molten metal
were considered as fundamental standards. Heroic observing programs 
were undertaken to tie the spectrum of Vega to
these laboratory sources. This accounts for the use of Vega as a
``primary'' spectral standard. In reality the primary metrology
standard was the blackbody source on Earth, of course.

The approach we suggest here exploits the availability of calibrated
photodiodes as NIST-traceable metrology standards, with spectral
response known at $10^{-3}$ level over the wavelengths relevant for
CCD instruments. This allows us to calibrate the photon sensitivity of
both imaging and dispersive apparatus as a function of wavelength,
using a tunable narrowband light source.  This technique allows us to
measure relative system throughput for multiple passbands, thereby
tying together the instrumental zeropoints across all filters.  With
this unambiguous system throughput data in hand, what remains is 1) to
determine a single overall absolute throughput (or ``effective
aperture'') for the system, common to all passbands and 2) ascertain
the optical transmission properties of the atmosphere.

\section{The arithmetic of broadband photometry}

Most of the imaging measurements in astronomy use array
detectors, with collecting optics and band-limiting optical filters,
to measure the flux from sources of interest.  Each pixel in the
detector is illuminated by both astronomical sources and background
radiation, termed ``sky'' in astronomical parlance. These different
sources have different spectral energy distributions (SEDs). The
signal $S_i$ from a source found in a pixel $i$ is then a sum
(over sources, plus sky) of integrals over wavelength,

\begin{equation}
S_i = \sum_{sources~j} \int \Fscr_j(\lambda)
   \times R_i(\lambda)\times A_i ~ d\lambda,
\label{eq:psignal}
\end{equation}

\noindent
where the $\Fscr_j$ are spectral photon distributions (SPDs, evaluated
above the atmosphere) of the sky and all sources present,
$R_i(\lambda)$ is the dimensionless system transmission of that
specific pixel, including atmosphere, optics, filter, and detector,
and $A_i$ is the effective collecting aperture of the telescope for
that pixel.  (We adopt the convention here that the units of $\Fscr$
are photons/nm/sec/cm$^2$, derived from the conventional spectral
energy distribution as $\Fscr = F_\lambda\;\lambda/hc$.)  We call the
product $R_i(\lambda)\times A_i$ the ``spectral aperture'' of a pixel,
it is that pixel's cross section for collection of light from a
celestial object impinging at the top of the atmosphere.

In general the transmissions $R_i(\lambda)$ are not constant from
pixel to pixel, varying even as a function of $\lambda$ (think of
non-uniform interference filters, water spots on detector
anti-reflection coatings, or the variations in internal interference giving rise to
fringing).  Conventional photometric analysis seeks to simplify this
by renormalizing the broadband $S_i$ to a signal $S$ with a known,
common spectral aperture $R(\lambda)\times A$, so that the resulting
signal does not depend on which pixel $i$ collected it.  This is
normally effected by division by a ``flatfield'', derived from a
source of light which is intended to have no variation from pixel to
pixel. It is obvious from Equation~\ref{eq:psignal}, however, that
there is no single flatfielding correction, even in principle, that
can be multiplicatively applied to the measured integral signal $S_i$
which can properly correct for the wavelength--dependent
$R_i(\lambda)$, for all possible SEDs.

For example, an undulating supernova SED, a smooth galaxy SED, and the
night sky emission lines might be all inextricably superposed on a
single pixel.  Because of internal fringing, a nearby pixel might have
larger QE at a prominent sky emission feature, hence report a greater
signal if the telescope were pointed to bring the feature onto this
pixel (i.e. a fringe maximum instead of minimum).  However, if the
night sky aurora diminishes, this ratio between the two pixels would
diminish, demonstrating that there is no unique flatfield scaling
factor between these two pixels.  Of course, sensible astronomers
recognize the additive contamination of sky emission and {\it
subtract} it, taking advantage of its temporal stability to expose a
pixel to a common night sky SED but an ensemble average of celestial
objects.  The supernova and galaxy SEDs, are similarly disentangled by
observing the galaxy when the supernova does not exist.  While this
illustrates important techniques of using time to improve differential
photometry, it is important to bear in mind that in principle
{\it each pixel} has its own sensitivity as a function of $\lambda$.

We argue below that it is possible to derive a common sensitivity
function for an array of pixels which will be accurate at the
sub-percent level for objects with SEDs which vary sufficiently slowly
with wavelength (rapidly varying SEDs such as night sky emission can
be handled by special procedures), but these considerations insist
that fluxes are best reported in such a ``natural system'' of the
apparatus, as simply the observed integrated flux over this common
passband.  Any SED-ignorant transformation to another photometric
system through some set of linear equations (known as ``color terms'')
is susceptible to systematic errors at the percent level or worse.
With SED in hand, of course, it is simple to integrate it over the
``natural system'' and another passband to derive the signal which
would have been collected in this passband.  A clear example arises in
supernova photometry, where using stars to derive transformation
coefficients between photometric systems produces systematic errors
due to differences in the SEDs of stars and supernovae.

Saha et al. \cite{Saha05} and Sung \& Bessell \cite{Sung00} provide clear discussions of the importance of fully understanding the interplay between passband variations and SED variations. 

\section{The arithmetic of light collection and detection}

The light from a point celestial object arrives at the telescope
nearly as a plane wave --- the distribution of arriving photons is
spatially uniform and essentially a delta function in angle.  The
telescope is a linear optical system which maps angles to positions
on the focal plane, and so there is a transfer function $H$ which 
describes how the intensity $\Phi$ at positions $\vec x$ in the 
focal plane is related to incoming photon flux $I$ at entrance
aperture location $\vec x^\prime$, pointing direction \bbeta~ of the
optical axis, and angle \balpha~ relative to the optical axis:
\begin{equation}
\Phi(\vec x, \bbeta, t, \lambda, P) = \int I(\vec x^\prime, \balpha,
\bbeta, t, \lambda, P) \; H(\vec x, \vec x^\prime, \balpha, \lambda, P)\; 
d^2\vec x^\prime \, d^2\balpha .
\end{equation}
The variables $\lambda$ and $P$ represent wavelength and polarization
(polarization is suppressed from now on).  This integral expression can in principle
be evaluated at any surface down the optical train. We find it most
convenient to consider the plane just above the first mechanical or
optical element, so that $I$ from a point source in the sky is an
unobscured plane wave of uniform surface intensity.

It is approximately true (and a goal of optical design) that $H$ be separable
\begin{equation}
H(\vec x, \vec x^\prime, \balpha, \lambda) \approx
\exp(i k \vec x \cdot \vec x^\prime / F)\;
\exp(-i k \balpha \cdot \vec x^\prime)\;
A(\vec x^\prime) \; \eta(\lambda),
\end{equation}
where $k=2\pi/\lambda$, $F$ is the focal length (not to be confused with 
the notation for SED), $A$ is the pupil 
transmission function (i.e. unity over the pupil, zero elsewhere), and
$\eta(\lambda)$ is the wavelength response of optics and filter, but
not including atmosphere or detector.
The response $\Phi$ at the focal plane is therefore the Fourier
transform of the (potentially complex, phase-error carrying) input $I$
at the entrance aperture, multiplied by a phase factor according to the
direction of incidence.  Ignoring the effects of diffraction, $H$ will
integrate to a non-zero response only when the argument of the
exponentials is zero,
\begin{equation}
\vec x = F \balpha,
\end{equation}
i.e. it maps angles to position, and picks up an amplitude factor of
$\sim\cos\alpha$.

Most contemporary astronomical detectors convert this incident photon
flux into photoelectrons where
\begin{equation}
S(\vec x) = \int \Phi(\vec x, \lambda)\; Q(\vec x, \lambda)\; d\lambda,
\end{equation}
where $Q$ is the detector's quantum efficiency, generally taken to
depend only weakly on sub-pixel location, angle of arrival, or
polarization.  The response from 
a given pixel is the area integral of $S(\vec x)$ over the extent of
that pixel, possibly convolved with diffusion within the detector or
cross-talk from other pixels which are being addressed at the same
time, run through an amplifier with a certain gain and linearity, and
digitized by an A/D with its own behavior which may not be ideal at
the 1\% level.  Understanding and suppressing imperfections in the
signal chain is a fairly well defined electrical engineering problem,
albeit one which merits careful consideration as we push for higher
performance from the apparatus.

The intensity which arrives at the entrance aperture is related to the
spectral photon distribution $\Fscr$ above the atmosphere
\begin{equation}
I(\lambda, \balpha, \bbeta, t) = \Fscr(\lambda, \balpha, \bbeta) \; T(\lambda,
\balpha, \bbeta, t),
\end{equation}
where $T$ is the atmospheric
transmission, which depends upon wavelength, direction, and time.  
The challenge of ground-based photometry is to convert
from the measured quantity, $S(\vec x)$, namely the distribution of
photoelectrons among the detector pixels, to $\Fscr(\lambda)$, the
photon spectra of the sources of interest.  This requires properly
understanding and correcting for 1) the transmission of the atmosphere
$T$, 2) the optical transmission properties of the apparatus, $H$, and
3) the detector quantum efficiency $Q$.

We advocate bundling the apparatus terms into the product $HQ$, which
we should strive to measure in a way that minimizes systematic
uncertainties, while considering the determination of the atmospheric
transmission $T$ as a distinct problem that merits its own dedicated
monitor.  We expect that the $HQ$ product will not depend on where the
telescope points ($\bbeta$) and it should vary slowly with time, so it
does not need to measured for every exposure.  We also have full
experimental access to the apparatus so we can measure $HQ$ to
arbitrary accuracy, 
at least in principle.  The $T$ term, however, should be measured for
each exposure, with recognition that it bears some dramatic spectral
features.

\section{Calibration of telescope and detector}

The availability of well characterized silicon photodiodes, with
sensitivity vs. wavelength curves that can be traced to fundamental
metrology standards \cite{Larason98}, provides us with the opportunity
of exploiting these devices to perform a relative calibration of
astronomical apparatus.  This reference detector can be used in
conjunction with a light source of arbitrary brightness to obtain a
throughput vs. wavelength curve for astronomical apparatus.

In principle, for each detector pixel we could create an appropriate
monochromatic plane wave of known intensity at the entrance aperture
of the telescope and thereby derive its spectral aperture function.
This would also tell us the extent to which $H$ does not integrate to
a delta function in angle \balpha~ and detector position $\vec x$,
e.g. scattered or diffracted light, and it duplicates the illumination
from celestial objects across the entrance aperture, so sensitivity in
$H$ to $\vec x^\prime$ is removed.  

In practice, collection of such measurements would be too time
consuming, so we endorse the normal procedure of filling the telescope
entrance aperture with light from a flat field screen and thereby
simultaneously filling the cone of angles which illuminates the entire
focal plane.  However, we advocate using {\it monochromatic} light and
monitoring the light level with a calibrated photodiode.  Obtaining a
series of flat fields at a discrete set of wavelengths and normalizing
each flatfield to the photon dose it received then provides us
directly with a data cube which is an approximation of each pixel's
spectral aperture.  
Another approach to this problem is described in \cite{Marshall05}.

This approximation fails to take account of breadth of 
$(F\balpha-\vec x)$ 
(``scattered light''), and we do not have a better suggestion than
to appeal to on-sky measurements to correct for this.  We hope that
the ``scattered light'' has the same spectral characteristics as the
direct transmission part of $H$ (i.e. if you paint your telescope
structure red you will need to measure the scattered light in each of
your filters), that it is spatially smooth, and temporally stable, so
the flatfield data cube can be multiplied by a smooth scaling function
(which ideally is not a function of wavelength) to convert it to the
accurate spectral aperture we seek.

This ``illumination correction'' function is normally derived by
rastering celestial objects across the detector during times that the
atmosphere transmission is thought to be extremely uniform and
constant.  This correction is normally found to contribute at the few
percent level, depending on the geometry of light scattering paths in
the telescope and optics as well as the illumination pattern from the
flatfield screen.  Manfroid \cite{Manfroid95} \cite{Manfroid96}, and Magnier and Cuillandre \cite{Magnier04} 
provide a thorough description of their experience in this regard. We obviously advocate a flatfield illumination
scheme which is extremely stable and uniform, and we next quantify how
uniform it must be to achieve sub-percent photometry.

\subsection{Coupling between flatfield illumination and photometry 
non-uniformities}

For the scheme outlined above, where a flatfield screen is used to
illuminate the entire focal plane, a number of effects can produce
systematic errors in throughput measurements. We have made
quantitative estimates of the corresponding requirements in
illumination.  Broadly speaking we can make errors in the spatial
uniformity of the density of photons crossing the entrance aperture
and we can make errors in their angular distribution.  We take as
given that an ``illumination correction'' will be performed; therefore
many non-uniformities which are not wavelength dependent will be
calibrated out and not contribute systematic error.

Observers ordinarily have to worry about matching the SED of the
flatfield illumination to that of celestial objects (perhaps a 3000K
incandescent filament with a color balance filter), but different
portions of the flatfield screen may have significantly different SEDs
(e.g. when illuminated by more than one flatfield lamp).  By creating
a data cube of normalized monochromatic flatfields we postpone this
issue until we collapse the cube to a mean flatfield.  This is
discussed below.

Spatial non-uniformities in illumination affect flatfield fidelity, but
the ratio between photometry uniformity (i.e. flatfield fidelity) and
spatial illumination uniformity is of order 5-10\%, i.e. a relatively crude
illumination density is tolerable.  The extent to which ``illumination
correction'' can mitigate these effects depends strongly on the
resolution of the illumination correction, but we assume that this
resolution is low.  Effects which couple spatial
non-uniformity to photometric non-uniformity include:

\begin{itemize}
\item{} A pixel corresponding to an angle \balpha~ receives
light which is offset on the flatfield screen by $\balpha d$, where
$d$ is the distance from the entrance aperture to the screen.  The
photometric uniformity is related to the integral illumination from
these offset but partially overlapping patches on the screen.  For the geometry
of the MOSAIC imager at CTIO, for example, we estimate that we could
tolerate large scale fractional surface brightness variation as large
as $\delta I/I\sim 15\%$.

\item{} The transfer function $H$ may have spatially
dependent terms (for example obscuration, dust on optics, degraded
mirror coatings, or vignetting).  Spatial non-uniformities in
flatfield illumination thereby translate to spatial non-uniformities
in detector illumination.  We estimate that the blurring effect of the
integral over $H$ reduce the latter to roughly 10\% of the former,
i.e. 10\% illumination variations can still yield 1\% photometry,
although this depends on details of the nature of the obscuration and
where it is.

\item{} Our scheme depends on a flux normalization from a
photodiode which does not exactly monitor the light seen by a typical
pixel.  We are then susceptible to normalization errors if this
mismatch is wavelength dependent.

\end{itemize}

The angular distribution within the entrance cone which
maps to the detector translates directly to photometric accuracy,
although the ``illumination correction'' procedure alluded to above
can in principle correct any non-uniform angular pattern which is
stable.  This distribution must therefore be uniform (or known) at the
sub-percent level for sub-percent photometry.

\begin{itemize}
\item{} Spatial non-uniformities of the flatfield screen
angular response are another case of spatial non-uniformity described
above.  Flatfield screens are often lit at a glancing angle,
and the scattering function between the illumination direction and the
(nearly) normal direction into the telescope can vary markedly at
extreme angles.

\item{} Spatial non-uniformities across the screen can
also cause bandpass errors.  We normally use interference filters in a
converging beam which undergo a wavelength shift which depends on the
angle of incidence:
\begin{equation}
\lambda(\theta) = \lambda_0 (1-\sin^2\theta/n_e^2)^{1/2} \sim 
                  \lambda_0 (1-\theta^2/8).
\end{equation}
For example, illumination of the CTIO 4-m flatfield screen by a
Lambertian source at the top of the prime focus has a radial intensity
fall-off across the screen which depopulates the outermost cone
angles.  At 700~nm, the average wavelength shift would be 1.2~nm, as
compared with 1.5~nm for a uniformly filled cone, so such a flatfield
passes through a filter which is 0.3~nm redder than that encountered
by a celestial object.

\item{} A ``thousand points of light'' flatfield screen
consisting of 1000 downward-pointing fibers distributed randomly
across the entrance aperture is not only spatially non-uniform, but
would require very careful injection of light since fibers
``remember'' the angular distribution of light fed into them for a
great distance.
\end{itemize}

The angular distribution outside the detector entrance cone
will create scattered light which may differ from that from the sky. 
The desire for a very uniform angular distribution usually leads to an
extremely wide illumination pattern, generally the full $2\pi$
steradians.  Effects which affect the flatfield include:

\begin{itemize}
\item{} The angular distribution of scattered light from celestial
objects is usually limited by the dome aperture, whereas a flatfield
screen radiates at all angles, giving rise to quite different
scattering contributions.  Attempts to mitigage this effect by
limiting the spatial size of the screen to the entrance aperture
have very little utility in a modern dome which closely fits the
telescope.  The ``illumination correction'' is essential for flatfield
screens which radiate significantly outside of the entrance cone.

\item{} Observers are all familiar with the dramatic smears of
scattered light occuring when moonlight illuminates the telescope 
structure.  Similar effects occur from sources of light in the dome
other than the flatfield screen, for example daylight leaking in
through cracks or instrumentation LEDs.  We argue below that it is
vital that flatfield illumination be shutterable, so that a
``flatfield'' is the difference between the lit and dark screen.
This differential technique will be effective to the extent that temporal variation in light leaks in the dome are slow
compared to the on-off sequencing time. Cloudless days are
therefore optimal for obtaining these calibration data.
\end{itemize}

In summary we desire the angular uniformity of our flatfield source to
match the photometric accuracy desired (or at least stable at that
level), but the spatial uniformity can be $\sim10\times$ less uniform
because obscuration couples in only at the 5--10\% level or so.
Scattered light is likely to be a significant contributor no matter
what we do, implying that application of an ``illumination
correction'' function is required.  We therefore want to try to
maxmize the smoothness of the scattered light contribution and
minimize its wavelength dependence.

\subsection{Implementing this Calibration Scheme}

In order to characterize the system throughput we need to provide
three components: a source of monochromatic light, a means of injecting
it into the telescope aperture, and a way to measure its
intensity.

\subsubsection{Monochromatic light source}

An $0.25^{\prime\prime}$ pixel requires some $10^4$ photons for a
flatfield, but subtends only $2\times10^{-13}$ of $2\pi$ steradians,
so at 20\% efficiency a light source must provide 0.1~joule to a
$2\pi$ steradian injection system.  While a monochromator can do this,
the tradeoff between spectral purity (we need $\sim$1~nm to see
fringing) and exposure time (acquisition of a 500 wavelength data cube
through $N$ filters is some $1000N$ exposures) is very challenging unless
we can deliver the light into much less than $2\pi$ steradians.

Another scheme is a pulsed tunable laser (the Vibrant by Opotek is one
example) which offers many advantages. These devices can deliver
tens to hundreds of mW of power into an optical fiber, 
tunable across the wavelengths of interest. The short coherence length
of the light avoids any speckle effects. We have installed such a unit
at CTIO and are pleased with the results \cite{Stubbs05}.

\subsubsection{The Flatfield Screen}

As outlined above, we wish to provide a full-pupil illumination source
that has surface brightness uniform to 10\%.  Most flat field systems
are reflective: one or more sources project light towards a screen
that then reflects it back into the telescope system.  A simple
reimaging system with a single on-axis source will exhibit a radial intensity
dependence of roughly $cos^4(\theta)$ where $\theta$ is the angle from
the telescope axis to the screen element $dA$, measured at the source.
Minimizing this requires a large standoff distance, in order to reduce
the angle $\theta$.

Modern telescope enclosures are engineered so as to minimize the
enclosed volume, and the top end of the telescope is often
precariously close to the inner wall of the enclosure. This lack of 
stand-off distance makes it very difficult to achieve unobscured, uniform
surface brightness illumination with a reflective scheme. 

We are in the
process of developing a uniform back-lit flatfield screen. An optical 
fiber is placed between a reflective 
backing and a diffusing screen. The optical fiber is engineered to 
produce substantial light leakage along its length, and the layout of the
fiber in the sandwich is designed to produce uniform surface brightness over the 
surface of the screen, compensating for the variation in brightness along the 
fiber's length. 
Tests we have carried out on prototypes of this approach 
are very promising, and we intend to implement this scheme for both the 
Pan-Starrs and LSST systems. 

\subsubsection{The illumination monitor}

In order to map out the spectral response of the telescope we must have a
means of measuring the light intensity from the flatfield
screen.  This can be readily achieved with a calibrated photodiode.

Ideally the photodiode would measure the light emanating from the
entire extent of the flatfield screen within the same solid angle
destined for the focal plane.  This may not practical, however, so the
photodiode's view of the flatfield may not sample the entire spatial
extent, and/or it may see a larger solid angle than does the focal
plane.  Obviously an effective design depends on the spatial and
angular uniformity of the flatfield emission, so we advocate analysis
and experimentation to find the best compromise for photodiode
illumination, given the dominant systematic errors of a given
flatfield implementation.  We have had good success in using the
photodiode in a simple pinhole camera configuration, monitoring the
light emanating from the entire surface of the flatfield screen.

\subsubsection{A focal plane monitor?}

We note that placing another calibrated photodiode in the focal plane
is valuable for separating the throughput of the telescope from the QE
of the detector.  Aging aluminum or filters can thereby be
distinguished from CCDs onto which outgassed substances are
condensing.  To the extent that the CCD QE is stable (which is likely
to be the case over timescales of months), this also provides a
measure of the gain of the CCD signal chain.

\section{Measuring the optical properties of the atmosphere}

A trio of phenomena dominate the transmission properties of the
atmosphere \cite{Slater80} \cite{Houghton77}.   The physical processes
\footnote{Scintillation can also produce photometric 
variation for systems with aperture $D$ comparable to the 
atmospheric Fried parameter $r_0$. We will assume all integrations
are long enough to suppress this effect to below the 1\% level.}
 that
determine the atmospheric transmission $T(\lambda)$ include:

\begin{enumerate}

\item{} Rayleigh scattering from the atoms and molecules of
atmospheric gases,

\item{} Small particle scattering from aerosols suspended in the
atmosphere, and  

\item{} Molecular absorption, in particular by $O_2$, $O_3$, and $H_2O$. 

\end{enumerate}

In addition to the imperfect transmissive properties of the Earth's
atmosphere, at wavelengths longer than 7500 \AA ~the atmosphere is a
significant source of narrowband emission, mostly from OH molecular
transitions in the ionosphere.
 
The atmospheric transmission can be a complicated function of airmass,
particularly at wavelengths where molecular absorption is a major
factor. These absorption features are responsible for the ``telluric''
absorption lines in ground-based spectra, and as they approach
saturation the airmass dependence is a complicated function of airmass
\cite{Adelman96}, \cite{Houghton77}, \cite{Guiterrez82}, \cite{Schuster01}.  
Assuming broadband transmission with a simple scaling with airmass is
inappropriate for precision compensation for atmospheric absorption, a
fact that is fully appreciated among infrared astronomers
\cite{Manduca79}.

With the atmospheric transmission factored out as a separate
measurement problem, we can use a dedicated piece of apparatus to
monitor celestial sources in order to measure $T(\lambda)$. This
can be done with multiband photometry, with a dispersive instrument, or both.
We propose making measurements of atmospheric transmission at the
time, and along the same line of sight, as the primary instrument is
being used.  This will require separate stand--alone apparatus.

The Pan-Starrs project intends to use a combination of a 
wide-field single-chip imager to monitor transparency across
the entire FOV of the imager, in 
conjunction with a spectrograph to monitor the wavelength
dependence of atmospheric throughput. The wide-field imaging
aspect has been in routine use at CFHT for some time \cite{Cuillandre02}. Here we will focus on the dispersive measurement. 
 
Fortunately, much is known about the optical properties of the
atmosphere \cite{Slater80}, so extracting a parametric description of
its transmission qualities should not be difficult. Unfortunately we
know surprisingly little about the angular and temporal correlations
of atmospheric emission and transmission, so the conservative approach
is to ensure that the $T$ system is keeping pace with the main
telescope.

One advantage to using a dispersive instrument to monitor
atmospheric transmission is that the {\it emission} spectrum of the
sky is also then available. Since many astronomical sources of
interest are significantly fainter than the sky, in should in
principle (knowing the system response $R(\lambda)$) be possible to
estimate the number of sky counts each pixel $i$ receives, and
explicitly subtract this from the sum in equation (1).

We intend to use a modest telescope and a commercial fiber-fed low
dispersion spectrograph to monitor appropriate sources within or close
to the fields being observed with the Blanco 4m.  We will explore how
to best extract a parametric description of the 3 processes of
concern, namely Rayleigh scattering, aerosol scattering, and molecular
absorption.  We intend to use these parameters as input to a
comprehensive computer model of the atmosphere (such as Modtran), in
order to allow us to correct for molecular absorption features that
are well below our spectral resolution.

Initial progress along these lines is described by \citep{Granett05}. 

\section{Image Processing Implications and Opportunities}

\subsection{Mean spectral response and flatfield construction}

Ignoring issues such as sensitivity to polarization and photon arrival
direction, the construction of a {\it monochromatic} flatfield from
our flatfield datacube is simple in principle, although the removal
of the scattered light component which might differ in intensity or
SED from celestial light is an experimental challenge.  We normally do
not observe in monochromatic light, however, but through relatively
broad filters with $\Delta\lambda/\lambda\sim0.2$, and we have argued
above that it is therefore impossible to correct the flux ratios
reported by different pixels to a consistent answer for all celestial
sources.  On the other hand, the mathematically consistent approach of
reporting the observed pixel responses along with a data cube of
response functions for each pixel (effectively the
illumination-corrected, normalized flatfield data cube) is not
satisfactory either.  We do recognize that there are celestial sources
such as planetary nebulae or HII regions whose emission is essentially
a delta function in wavelength for which this is the only approach
which will work.  (Even then caution is necessary: we have alluded to
the wavelength shift of a flatfield arising from non-uniform screen
illumination that could cause large error for an emission line which
falls right at the edge of a bandpass.)

One need only glance at the image of a monochromatic planewave to
appreciate that there is indeed a very substantial variability in the
pixel-to-pixel response.  Fringing in the near IR causes a response
oscillation in $Q$ every $d\lambda = \lambda^2/(2nw)$ or equivalently
a change in thickness by $\lambda/(2n)$, where $w$ is the thickness of
the detector and $n\sim3.5$ is the index of refraction of silicon.
Diffraction around dust particles also creates a wavelength-dependent
response of $H$ at all wavelengths.  These effects are reduced by a
fast f/ratio and a broad cone of incoming light, but they are usually
present at the few percent level at least, and a thin device with an
AR coating not optimized for the IR can fringe at the 10\% level or more.

The response to celestial sources is the integral over wavelength of
the pixel response function $R_i$ times the celestial SPD.  If we wish
to use this to infer what a system with a {\it mean} response $R$
would be to this SPD, we will make a fractional error which is

\begin{equation}
{\delta S \over S} = {\int \Fscr(\lambda) \; (R(\lambda)-R_i(\lambda))
       \; d\lambda \over \int \Fscr(\lambda) \; R_i(\lambda)\; d\lambda}.
\label{eq:fracerr}
\end{equation}

\noindent
Obviously the error will be large for objects where changes in
$\Fscr(\lambda)$ coincide with deviations $R_i(\lambda)-R(\lambda)$,
integrated over the bandpass.  Conversely, if we can keep
$R_i(\lambda)-R(\lambda)$ relatively small, $\Fscr(\lambda)$
relatively smooth, and the bandpass is relatively wide, the error will
be small.  These three factors can be traded off against one
another: if we were willing to keep track of each pixel's
$R_i(\lambda)$ there would be no error; trying to go from a pixel's
$R_i$ to some grossly different bandpass $R_{std}$ will
incur larger error; objects with smooth $\Fscr(\lambda)$ will have
more accurate results than those which vary, etc.  Integration of
equation~(\ref{eq:fracerr}) for ``reasonable'' SPD and detector
response over a ``reasonably'' wide bandpass indeed yields errors
which are less than 1 percent.  

We consider a ``reasonably'' smooth SPD to have no more than isolated
20\% variation per nm (e.g. H\&K break or molecular band edges), and
otherwise 5\% variation per nm (e.g. supernovae).  A ``reasonable''
ratio $R_i/R$ between pixel response and mean response
does not exceed a 2\%/nm derivative at bandpass edge, a slope
of 2\% across a bandpass, and oscillations of 10\% with period 2.5~nm
in regions redder than 750~nm, where sky emission lines become
important.  A ``reasonably'' wide bandpass is $\Delta \lambda/\lambda
\sim 0.2$.  We regard objects which do not meet this criterion to be
``photometric challenges'' and analysis seeking sub-percent photometry
should pay 
attention to the individual $R_i$ lest an unfortunate coincidence of
emission or band edge with response oscillation cause large
photometric error.  We regard pixels which do not meet this criterion
to be ``photometrically challenged'' and should be flagged as such.
We assert that it is sufficient, for most applications of interest to
astronomers, to construct a single system response function
$R(\lambda)$ per passband, along with a single 2-d flatfield image
that corrects for pixel-to-pixel variations and illumination
nonuniformities, appropriately weighted across the passband.  Both of
these can be extracted from the stack of monochromatic flatfield
images.

We advocate applying the multi-pronged technique used by most astronomers: {\it
subtraction} removal of sources with violently varying SPDs (emission
lines), {\it division} by the flatfield data cube which has been
integrated over a mean bandpass or a flatfield obtained from a
continuum source, reporting a flux value whose collection pixel
signature has been suppressed to below the 1\% level, and supplying a
mean response function vs. wavelength for the entire array.

The image processing stages that we consider appropriate, using the
approach described here, include:

\begin{itemize}
\item{} Begin by correcting the raw flatfield data cube, wavelength by
wavelength: 
\begin{itemize}
\item{} Remove ambient light contamination by subtracting source on/off pairs.
\item{} Normalize each flat to a uniform photon dose by dividing by the 
integrated signal seen by the calibrated photodiode, after accounting for 
its photon detection efficiency.  Correction for a
spatially non-uniform shutter response is appropriate here.
\item{} Divide out large-scale unwanted artifacts, such as edge
dimming from the Jacobian of the distortion (we want to preserve
fluxes after division by flatfield), but leaving in desired
features such as edge dimming from vignetting.
\item{} Calculate the illumination correction, by
integrating a nominal bandpass flatfield (described below), dividing
it into images of a star field taken under photometric conditions at
many different offsets, measuring the fluxes and using all ratios for
each star to estimate a model (e.g. polynomial or coarse grid) for the
illumination correction.
\item{} Multiply all the normalized wavelength slices by the illumination
correction model.
\end{itemize}

\item{} Construct a mean response function $R(\lambda)$ for all pixels
by averaging the responses $R_i(\lambda)$ of individual pixels over
the array.  (Each $R_i$ should be scaled to the same value first, we are not interested in wavelength independent dimming from effects such
as vignetting or distortion.)

%

\item{} Verify that differences between the mean response and that of
any pixel,  $R_i/R-1$,  does not exceed a 2\%/nm derivative
at bandpass edge, a slope of 2\% across a bandpass, and oscillations
of 10\% with period 2.5~nm in regions redder than 750~nm, where sky
emission lines become important. 

%
%
%

\item{} Verify that the integration of source SPDs of interest, over a
bandpass of 
width 100~nm, have location-dependent 
errors relative to the mean bandpass of $<1$\%.  Detectors which are
worse or SPDs which are more variable will have larger non-uniformity.
(It is possible to construct counterexamples of allowed SPDs and
responses which will not meet the resulting 1\% criterion.)

\item{} There is a degree of freedom in how we construct our
2-d flatfield, from the normalized monochromatic flats.  
Celestial objects approximate blackbodies between 1000~K
and infinite temperature, hence have SPDs (photon rate per nm) which
go approximately as $\lambda^{+20}$ (Wien exponential) to
$\lambda^{+0}$ (4500~K) to $\lambda^{-3}$ (Rayleigh-Jeans tail) across
the optical region of the spectrum.  Across a broad filter bandpass of
$\Delta\lambda/\lambda\sim0.2$ this amounts to a difference of 40\%.
More relevant, our requirements on the constancy of $R_i/R-1$ will
give us 1\% photometry for any of these SPDs, but it is important 
to recognize that our photometric accuracy is in the {\it
instrumental} system with mean response function $R(\lambda)$.

\item{} To the extent the flatfield matches the celestial SPD, the
pixel-to-pixel variation will be removed perfectly.  We therefore
should select some intermediate weighting function, e.g. none
($\lambda^{+0}$) or filament temperature 2800~K ($\lambda^{+3}$), and
integrate the flatfield datacube with that weight.  This is then
reflected in our mean, dimensionless response function.
Another option is to weight the monochromatic flats so as to 
correspond to a particular calibration spectrum of interest, 
such as AB magnitudes ($\lambda^{-1}$), although that doesn't
automatically confer additional accuracy in AB magnitudes.

%
%
%
%

\item{} It is acceptable to illuminate the flatfield screen with white
light (and thereby do an analog integration of the flatfield datacube)
but the division of image by flatfield impresses the SPD of the
flatfield on the data, so the flatfield SPD {\it is} a component of
the system response $R$ and must be measured.  Apart from the
advantages of measuring the system spectral response, we regard
measurement of a white light SPD and keeping it stable to be difficult
enough that we believe that it is preferable to accumulate and
integrate the monochromatic data cube.

\item{} If we were to estimate a flux in a ``standard'' filter which differs
from our actual bandpass without any correction for the source's SPD
we could clearly make large errors to the extent that the bandpasses
differ.  However, the traditional method of using ``color terms'' to
estimate an ideal response from an observed one is clearly an
approximation based on a linear (or sometimes non-linear) fit to a
one-parameter family of SPDs, usually stars or blackbodies.  These
cannot be used for 1\% photometry of SPDs which differ by more than
10\% from the SPD which generated them.  E.g. a set of color terms
created from stellar SPDs are adequate for 1\% photometry of most stars, possibly for galaxies, but {\it not} for supernovae.

\item{} It is possible to do a posteriori improvement of photometry if
we can estimate the SPD of an object of interest, for example, a
supernova at a given age and redshift.  We can then return to the
datacube of pixel response functions and integrate the pixel response
and mean response against both the supernova SPD and the chosen
flatfield SPD, and use the ratios to correct the flux from the
flattened image to that which would have been reported by pixels which
follow the mean response exactly.

\end{itemize}

\subsection{Explicit Sky Subtraction?}

One could imagine determining the sky emission spectrum using the data
from a dedicated dispersive sky monitor and a model for OH
emission. The contribution of the sky spectrum to the counts in each
pixel could then be subtracted using the monochromatic flatfield
stack.  The fringe pattern that arises from strong sky lines
illuminating a detector with a spatial variation in sensitivity is
normally very stable in character but can be quite variable in
amplitude. It remains to be seen whether the sky emission data will
provide a fringing correction that is superior to that obtained from
stacking images and subtracting an appropriately scaled fringe frame,
where the amplitude of fringing in each image must to be determined
before the master fringe frame can be scaled and subtracted.

\subsection{Photometric measurements}

We can hope to have reasonably accurate absolute photometry if we know
that the sky has a well behaved transparency, for example if the
conditions are ``photometric'' as judged by the sky monitor.  However,
photometry at the 1\% level is unlikely to be possible without
recourse to celestial photometric standards.  The present situation of
a handful of calibrated photometric standard stars is currently
improving via SDSS to photometrically calibrated stars found in every
field of view.  

The advent of the astrometric-photometric survey of Pan-STARRS which
should begin in 2006 will improve the situation yet more.  This will
extend photometric measurements to the 1\% level in a very well
characterized set of filters to the entire sky visible from Hawaii.
The accurate characterization of the Pan-STARRS bandpasses and
knowledge of the standard object's SPDs should permit use of these
object's calibration to bring any observation to an accurate value for
physical flux integrated over the bandpass in use.

As detectors become larger and photometry requirements become more
demanding, it becomes necessary to regard the sky transparency as a
variable across the field of view.  This can potentially be gauged
from the sky monitor, but a more likely procedure will be to use the
very high density of calibrated objects to map the variation in
transparency.

\section{Considerations for the design of new apparatus}

Many aspects which limit the precision of photometry are legacies of
the days before electronics, computers, and large scale detectors,
such as the use of a single ``exposure time'' instead of a spatially
variable shutter timing function.  Some improvements can be made with
a few new calibration procedures and subroutines, such as linearity
corrections.  Others require new apparatus.

\subsection{Signal chain electronics: accuracy and stability}

No digital data will be better than the analog electronics which
produces it.  A CCD output MOSFET dissipates about 10~mW which is
enough power to cause local warming.  A large signal can change this
power by 10\%, and a CCD amplifier can be sensitive enough to
temperature that this changes its gain.  This would show up as
non-linearity (which might be different for a bright star than a
flatfield), but of course this non-linearity could be sensitive to the
overall temperature of the CCD.  The near-IR QE of CCDs is very
sensitive to temperature as well, as much as 1\%/K at 1~$\mu$m.  This
strongly suggests that CCD temperatures be regulated to no worse than 1~K.

CCD amplifiers are typically non-linear near the 1\% level, and the
full well of the CCD pixels is often not reached because the amplifier
has become so non-linear that the data are useless.  There are many
satisfactory techniques to measure the linearity of a CCD amplifier
and we recommend that this (a) be done on a regular basis, (b) a
correction be routinely applied to the data stream, and (c) the
results be available as meta-data.  If the linearity changes with time
it is a sign that that amplifier and signal chain should not be
trusted for accurate photometry and it is important to know this.

The gain of an amplifier and signal chain is a related parameter.  In
principle it is removed by flatfield division, but in practice the
gain of an ailing CCD amplifier or signal chain A/D can also change
rapidly.  There are also techniques to monitor this gain at the few
percent level and again, it should be done regularly and available as
meta-data so that a misbehaving channel is flagged before it is
used to do high precision photometry.

\subsection{Filter Passband Width}

In order to ensure that pixel-to-pixel variations in $R_i(\lambda)$ do
not produce spurious results, it is important that the optical
passbands be wide enough to encompass many periods of sinusoidal
response variation with wavelength. The source SED plays a role here,
of course. In general we estimate that passbands of 
$\Delta\lambda/\lambda~\sim0.2$ should be sufficient to keep
position-dependent flux errors below 1\% for most sources of interest.
 
\subsection{Shutter timing}

Many observers pay attention to the fact that shutter timing is not
accurately reported by the instrument, and that shutter timing depends
on position in the focal plane; others do not.  Mechanical shutters
typically have non-uniformities at the 20~msec level, which translates
to a 1\% photometric error for a 2~sec exposure.  Of course this is a
common exposure for flatfields and bright photometric standards.  It
is straightforward to determine empirically the shutter timing
function over the detector by comparing short and long exposures if
the shutter is reproducible, but if the shutter's performance depends
on orientation or temperature this will not work.  We strongly advocate
that (a) shutters be equipped with sensors to report the shadow
trajectory across the detector to an accuracy of 0.1~msec, and (b)
this trajectory be coded into an exposure's metadata and
reduction pipelines be equipped to use a ``shutter timing function''
for reductions rather than a single exposure time. 

A recent example of shutter timing issues apparently producing 
systematic discrepancies in photometry is described in \cite{Stetson05}.

\subsection{Designing for calibration ease}

Pan-Starrs is an ambitious system nearing deployment; LSST and SNAP
are in the design phase.  For many applications of the data that these
systems will produce, systematics are likely to limit science. This
implies that the calibration aspects should be afforded a high
priority, including: (1) building in a good uniform flat field screen,
properly baffled, (2) designing for fast readout so that acquisition
of flatfield data is efficient, (3) allowing for more sophisticated
flatfielding and calibration processing in the reduction pipelines.

\section{Conclusions and Next Steps}

A different approach to the challenge of ground-based photometry, taking 
advantage of well-calibrated detectors as metrology standards, could
well allow measurements with relative flux uncertainties well below
the one percent level. An important ingredient is measuring directly
the optical properties of the atmosphere.

It is our intention to pursue these ideas as part of the ESSENCE
supernova survey, as well as in the Pan-Starrs and LSST projects.  We
are currently building and deploying equipment for the CTIO Blanco 4-m
telescope and 8k mosaic as part of the ESSENCE program.  We also
expect to implement most of these ideas on the first Pan-STARRS
telescope, scheduled for first light in mid-2006.

\section{Acknowledgments}

We are grateful to Nick Suntzeff for his patience during extensive
conversations about photometry, and for his advice, insight and
encouragement. We have also benefitted from conversations with Kem
Cook, Doug Welch, Jim Gunn, Michael Strauss, Tim Abbott, David
Schlegel, Robert Lupton, Chris Smith, Daniel Sherman, and the members
of the ESSENCE supernova survey. We are also grateful to Wesley Traub
for his insights on atmospheric modeling. 
The assistance of Eli Margalith of Opotek was essential to our enterprise, 
and we are most grateful for his help. 
This work was motivated by
the desire to achieve ground-based photometry with fractional
uncertainties at the percent level, in connection with the ESSENCE
supernova survey, and in anticipation of needs of the Pan-Starrs and
LSST systems. The ESSENCE project is supported by the National Science
Foundation under grant AST-0443378.
Our work on calibration considerations is also supported by the LSST Corporation.  We also thank the referee, Dr. Mike Bessell, for his insightful and constructive comments.

\clearpage

\end{document}